\documentclass[prl,twocolumn,showpacs]{revtex4}
\usepackage{graphicx}
\usepackage{amsmath}
\begin{document}
%
\title{Gas Enrichment at Liquid-Wall Interfaces}
\author{Stephan~M.~Dammer and Detlef~Lohse}
\affiliation{Department of Applied Physics, University of Twente, 7500
  AE Enschede, The Netherlands}
\date{\today}
\begin{abstract}Molecular dynamics simulations of Lennard-Jones systems are performed to study
  the effects of dissolved gas on liquid-wall and liquid-gas interfaces. Gas
  enrichment at walls is observed which for hydrophobic walls can exceed more
  than two orders of magnitude when compared to the gas density in the bulk
  liquid. As a consequence, the liquid structure close to the wall is
  considerably modified, leading to an enhanced wall slip. At liquid-gas
  interfaces gas enrichment is found which reduces the surface tension.\end{abstract}
\maketitle
The precise determination of the hydrodynamic boundary condition, {\em slip}
vs.~{\em no-slip}, is currently a matter of
active debate. A growing number of studies,
experiments~\cite{lauga,granickbizonne,bonaccurso} as well as
simulations~\cite{bizonneEPJE,barrat,priezjev,lichter,andrienko}, strongly indicate that
the classical no-slip condition, a more than 200 year old
dogma, is violated. Though it is difficult to
identify clear trends, the investigations suggest that increasing hydrophobicity
and an increasing amount of dissolved gas in the liquid favor larger slip. 
Note, however, that slip has been reported for hydrophilic surfaces as well~\cite{bonaccurso}.

Despite many investigations, slippage behavior and its origin are far from being understood.
A possible cause~\cite{lauga,degennes} is the presence of so-called {\em
  surface nanobubbles}, i.e., nanoscale bubbles located on a solid surface
that is immersed in liquid. Many recent experiments support the notion of surface
nanobubbles, in particular atomic force microscopy
measurements~\cite{nanobubbles_AFM}, but also other
techniques~\cite{nanobubbles_other}. Similar to the
trends for wall slip, hydrophobicity and dissolved gas favor nanobubbles. 
For gas-saturated liquid nanobubbles are found on hydrophobic surfaces,
whereas usually nanobubbles are not
observed for hydrophilic {\em and/or} degassed liquid, suggesting gas- rather than
vapor bubbles. In spite of growing experimental evidence for their existence,
it is unclear how and why they form and why they are apparently stable. 

Other examples for the interplay between hydrophobic interfaces and dissolved
gases are colloidal suspensions and emulsions~\cite{alfridssonmaeda}, where
the stability is considerably influenced by the
presence of dissolved gases. Moreover, recent neutron reflectivity measurements~\cite{doshi} reveal
a dependence of the width of the hydrophobic
wall-water interface on the amount and type of dissolved
gas.

Though the above mentioned experiments clearly demonstrate the importance of
dissolved gases for the hydrophobic wall-liquid interface, a profound
understanding is still lacking. Molecular dynamics simulations are a promising
approach to address this
issue. However, previous simulations, for instance of
slippage~\cite{bizonneEPJE,barrat,priezjev,lichter}, were
restricted to pure liquids without dissolved gases.
How do gases effect liquid-wall interfaces? How do the effects change with
hydrophobicity or for different gases? Is wall slip enhanced? It is
the aim of this Letter to address these issues by means of molecular dynamics
simulations. Control parameters are the {\em amount of dissolved gas}, the
{\em hydrophobicity of the wall}, and the {\em type of gas}. Liquid-gas interfaces, which serve as reference
systems and are important in their own right~\cite{lubetkin}, are studied as well. 

Simulations are performed for fixed particle number, volume and temperature
$T{=}300\,{\rm K}$ using the GROMACS code~\cite{gromacs}. Periodic boundary conditions
(p.b.c.) are applied in $x,y-$ and $z$-direction. Three different particle
species (liquid/gas/wall) with mass $m{=}20\,{\rm amu}$ are simulated. Liquid and wall
particles have the same molecular diameter $\sigma{=}0.34\,{\rm nm}$. Particles interact via
Lennard-Jones (LJ) 6-12 potentials with a cutoff $r_c{=}5\sigma$, which is
larger than the value $2.5\sigma$ usually applied for bulk liquids,
in order to account for inhomogeneities at interfaces. The energy scale
$\epsilon_{ll}$ for liquid-liquid interactions is fixed to
$\epsilon_{ll}{\approx} 1.2k_BT$ with Boltzmann's constant $k_B$. 
To model an inert gas without wall affinity the energy scales for
gas-gas and gas-wall interactions are $\epsilon_{gg}{=}\epsilon_{gw}{\approx}
0.4k_BT$, which is close to $\epsilon_{gg}$ of Argon. The temperature $T$ is
below (above) the critical temperature $T_c$ of the
liquid (gas) particles~\cite{footnote,brovchenko}. The time step is
$dt{=}0.005\tau$ with the characteristic
time $\tau{=}\sigma\sqrt{m/\epsilon_{ll}}{\approx} 0.9\,{\rm ps}$. During production runs,
the simulations are weakly coupled to a heat bath using the Berendsen
thermostat~\cite{berendsen} with a relaxation time $\tau _T{=}10\tau$. A
perfectly stiff wall is simulated by solid particles that
are frozen on a fcc-lattice with density $\rho_w{\approx} 0.96\sigma ^{-3}$. 
The center of mass velocity is removed, apart from the flow simulations.

Four {\em microscopic control parameters} (i)-(iv) are tuned which change the
properties of interest. 
To simulate different gases (i) the energy scale
$\epsilon_{gl}$ for gas-liquid interactions is varied, as well as (ii) the
molecular diameter $\sigma_g$ of the gas particles
($\sigma_{gl}{=}\sigma_{gw}{=}0.5(\sigma_g{+}\sigma)$ is applied). 
Expressing $\epsilon_{gl}$ and
$\sigma_g$ in terms of $k_BT$ and $\sigma$, respectively, the combinations $(\epsilon_{gl},\sigma_g){=}(0.4,1),$
$(0.692,1.47),$ $(0.692,1.62),$ and $(0.712,1.62)$ are
studied, which are denoted as gas types $(A){-}(D)$. To identify effects due
to the gas (iii) the number of gas particles $N_{g}$
is changed from $N_{g}{=}0$ (pure liquid) to the finite value $N_g{=}228$. 
The hydrophobicity of the wall is varied by (iv) the ratio
$\epsilon_{lw}/\epsilon_{ll}$ with the scale $\epsilon_{lw}$ for liquid-wall
interactions, enabling simulations of hydrophilic and hydrophobic walls. These
{\em microscopic} parameters (i)-(iv) determine {\em macroscopic}
properties such as gas solubility, gas concentration, surface tension, and the
contact angle. 

\begin{figure}[t]
\includegraphics[width=77mm]{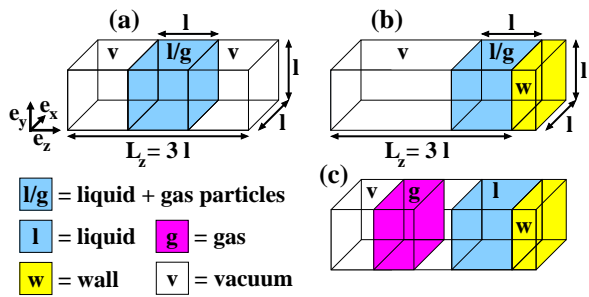}\vspace{0mm}\\
\caption{
\label{setup}
(color online) Starting configurations to study (a) liquid-gas
interfaces, and (b,c) liquid films at walls. Initially the particles are located on a
lattice, which 'melts' during equilibration, forming a liquid while the vacuum
is filled by a vapor-gas phase. Due to p.b.c. in (b,c) the wall
terminates the vapor-gas phase in $z$-direction. The scale is given by
$l{\approx} 16\sigma$.   
}
\end{figure}

\begin{figure}[t]
\includegraphics[width=85mm]{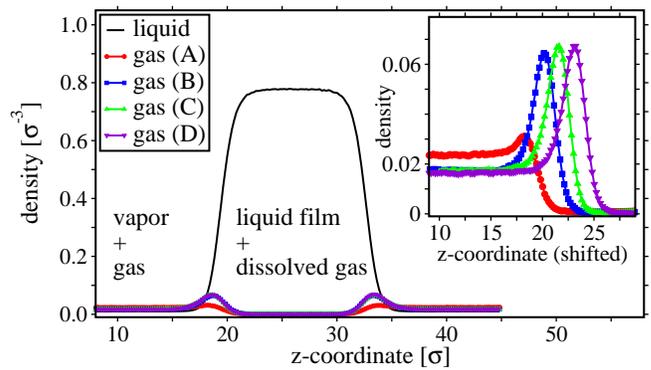}\vspace{0mm}
\caption{
\label{densfree}
(color online) Density profiles for liquid-gas mixtures at phase
coexistence. Gas enrichment at the interface can be observed (see inset). The
liquid profiles were similar for the different gases, wherefore only one is
displayed. Varying parameters for the gases $(A){-}(D)$ are
$(\epsilon_{gl}/\epsilon_{gg},\sigma_g/\sigma){=}(1,1),$ $(1.73,1.47),$
$(1.73,1.62),$ and $(1.78,1.62)$. 
}
\end{figure}

Before addressing liquid-gas mixtures at walls it is worth discussing
liquid-gas mixtures without walls. Initially, liquid and
gas particles are located on a lattice ('fluid cube')
in the center of a rectangular simulation box, Fig.~\ref{setup}(a). 
After an equilibration period ($9{\times} 10^6dt{\approx}40\,{\rm ns}$), which
consists of a series of subsequent microcanonical simulations at
$T{=}300\,{\rm K}$, the system is in a steady state with a liquid film
perpendicular to the $z$-axis, in coexistence with the vapor-gas phase. The
total number of particles is $N{=}2916$. Fig.~\ref{densfree} presents density
profiles obtained from time averaging
($10^6dt$ after equilibration). One can clearly observe an enrichment of gas in the interfacial
region, before the gas density falls off towards its value in the bulk
liquid (similar observations have been made for liquid-liquid
mixtures~\cite{salomons} with a much stronger attraction between different
particle species). A gas particle close to the interface experiences attractive forces
from particles in the vapor-gas phase as well as in the liquid film. Since the
density in the liquid film is much larger than in the vapor-gas phase, the
resulting force is directed towards the liquid film, which leads to the
nonmonotonous density profiles, even for gases with small gas-liquid
interactions as for $(A)$. Note that the amount of gas in the bulk of the
liquid is similar for all gases, $(6.25\pm 2.75){\times} 10^{-4}\sigma^{-3}$, though the importance of
different factors involved in the process of gas solution are expected to
differ for the gases $(A){-}(D)$. To illustrate this, consider the
energy scale $\epsilon_{gl}$ of $(D)$ which will
facilitate gas solution compared to (A) with its smaller value of
$\epsilon_{gl}$, but this is counteracted by the larger size of the $(D)$
particles.    

Does gas change the surface tension $\gamma$?
Experiments show that gases decrease $\gamma$ (which has
been proposed to be crucial for bubble nucleation), but the
reason was stated to be unknown~\cite{lubetkin}. Applying
the standard Kirkwood-Buff~\cite{rowlinson} formula to calculate $\gamma$, a decrease
of $\gamma$ due to the gas adsorption is found as well. Expressed in terms of
$\epsilon_{ll}/\sigma^2$, the average value of $\gamma$ for the liquid-vapor
interface ($N_g{=}0$) is $\gamma{\approx}0.74$ (in agreement to typical values
for LJ-fluids~\cite{salomons}), which is reduced to approximately
$(0.72,0.6,0.56,0.56)$ for $(A){-}(D)$. Additional
simulations show that the decrease of $\gamma$ is enhanced for increasing gas
pressure, just as in experiments. 

\begin{figure}[t]
  \vspace{-5mm}\hspace*{-1.1cm} \includegraphics[width=105mm]{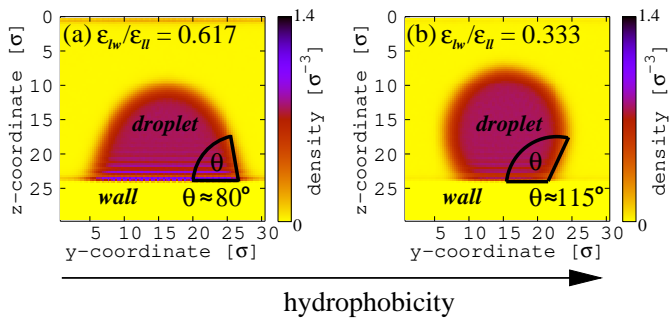}\vspace{0mm}
\caption{
\label{droplets}
(color online) Liquid density profiles of droplets at walls, to characterize
the walls. Tuning the {\em microscopic} attraction ratio $\epsilon_{lw}/\epsilon_{ll}$
(with fixed $\epsilon_{ll}$) results in a change of the {\em macroscopic}
observable $\theta$. Left, (a): hydrophilic wall,
$\epsilon_{lw}/\epsilon_{ll}{=}0.617$ with a measured contact angle
$\theta{\approx} 80^\circ$. Right, (b): hydrophobic wall,
$\epsilon_{lw}/\epsilon_{ll}{=}0.333$ with $\theta{\approx} 115^\circ$. 
The $x$-extension of the initial droplets ('fluid cube') equals the
$x$-extension of the simulation box, leading to (hemicylindrical) droplets,
translationally invariant in $x$-direction. 
}
\end{figure}

What changes in the presence of walls? Since the contact angle $\theta$ is of
vital importance, the walls are first characterized by simulations of
droplets at walls. Therefore, a 'fluid cube' composed
of liquid particles is initiated on a wall. After dynamical evolution
($2.5{\times}10^6dt$) the density profiles (obtained by $10^6dt$ time
averaging) allow to estimate $\theta$ numerically, see Fig.~\ref{droplets}. 
The trend of the obtained contact angle with the hydrophobicity parameter
$\epsilon_{lw}/\epsilon_{ll}$ is consistent with what one would obtain from
the rough estimate~\cite{barrat}
$\cos{\theta_L}{\approx}{-}{1}{+}2(\rho_w\epsilon_{lw})/(\rho_l\epsilon_{ll})$
(with densities $\rho_w$ and $\rho_l$ of wall and liquid) based on the Laplace
expression of surface energies~\cite{rowlinson}.  

\begin{figure*}[t]
\includegraphics[width=180mm]{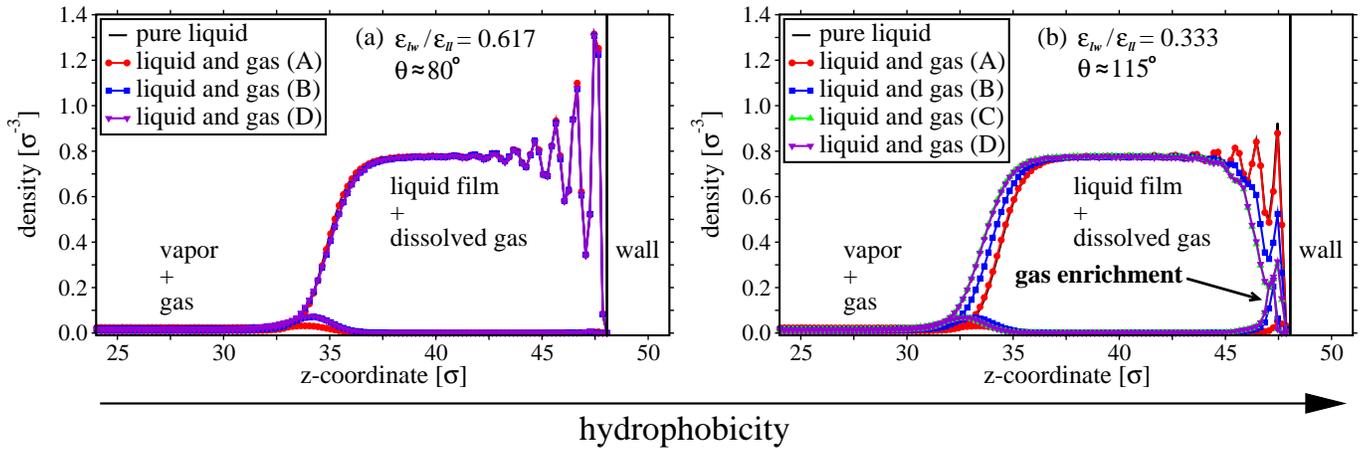}
\caption{
\label{densities}
(color online) Liquid and gas density profiles (same symbols for liquid
and gas) for liquid films in contact with walls. Left, (a): hydrophilic wall,
the liquid exhibits the usual layering which is not altered by the the
presence of the gases. Right, (b): hydrophobic wall, note the tremendous
increase of the gas density and the greatly diminished liquid density at the
wall.  Varying parameters for the gases $(A){-}(D)$ are
$(\epsilon_{gl}/\epsilon_{gg},\sigma_g/\sigma){=}(1,1),$ $(1.73,1.47),$
$(1.73,1.62),$ and $(1.78,1.62)$. 
} 
\end{figure*}

What is the molecular structure of liquids in contact with walls, in
particular in the presence of dissolved gas? With well controled walls in
place, we proceed to investigate this issue. Therefore a 'fluid cube' of
liquid and gas particles is initiated close to a wall, Fig.~\ref{setup}(b). The {\em effect of
  hydrophobicity} is studied by changing $\epsilon_{lw}/\epsilon_{ll}$, as
discussed above. In order to probe the {\em effect of dissolved gas} we
compare simulations with $N_g{=}0$ (pure liquid) to simulations with
$N_g{=}228$ for the gases $(A){-}(D)$. The number of liquid particles is
$N_l{=}2688$. After an equilibrating period ($9{\times}10^6dt$ and $12.4{\times}10^6dt$
for the hydrophilic and hydrophobic wall) the
system consists of a fluid film in phase coexistence on
one side and which is in contact with a wall on the other side. 
The left part of Fig.~\ref{densities} shows liquid and gas density profiles
($10^6dt$ time averaging) close to the hydrophilic wall. 
There is a tiny increase of the gas density at the wall, which is too small to
be observed on the scale of Fig.~\ref{densities}. The pure liquid exhibits
the usual layering~\cite{barrat,brovchenko,mugele} close to the wall, which is
hardly altered by the presence of the gas. 

How do the density profiles change for a hydrophobic instead of a hydrophilic wall? 
Fig.~\ref{densities}(b) shows liquid and gas density profiles close to the
hydrophobic wall. One immediately
observes a dramatic increase of the gas density in the vicinity
of the wall. For gas $(A)$ the density increases by a factor ${\approx} 50$
when compared to the gas density in the bulk liquid, and the gas enrichment is
even more than two orders of magnitude for the gases $(B){-}(D)$. Furthermore,
the liquid structure close to the wall is drastically
changed. The pure liquid exhibits layering, which is less pronounced than for the
hydrophilic case, Fig.~\ref{densities}(a). The liquid structure is
only slightly altered by gas $(A)$ but it is greatly
diminished by the presence of gases $(B){-}(D)$. The gas enrichment leads to a
considerable reduction of the liquid density in the
vicinity of the hydrophobic wall. We stress that for the hydrophobic wall for
all gases $\epsilon_{gw}{=}\epsilon_{lw}$, which shows that the gas enrichment is {\em
  not} caused by a strong gas-wall interaction. The gas enrichment is
associated with a reduced diffusion of the
gas perpendicular to the wall. Since the simulations leading to
Fig.~\ref{densities} started with a high gas concentration in the liquid,
Fig.~\ref{setup}(b), a proper equilibration
is a delicate issue (for gases $(B){-}(D)$). Therefore we confirmed the
results by simulations starting from a contrary initial
configuration, Fig.~\ref{setup}(c). Here, the gas particles are initially
completely separated from the wall by the liquid film. The gas enrichment
obtained after $12.4{\times}10^6dt$ equilibration from
simulations initiated from Fig.~\ref{setup}(c) is $71\%$, $56\%$, and $92\%$
of the gas enrichment for the gases $(B){-}(D)$ depicted in
Fig.~\ref{densities}(b)~\cite{footnote2}. Hence, the tremendous gas enrichment
as well as the considerable reduction of the liquid density at the wall are
reproduced even in simulations starting from the configuration
Fig.~\ref{setup}(c). 

The width of the region of
gas enrichment is of the order of $\sigma$, similar to the observations
in~\cite{doshi}. Thus, we basically find a
monolayer of gas particles adsorbed at the
wall. Though the gas enrichment shows some resemblance of surface
nanobubbles it is still different from gas bubbles with heights of several
nanometers as experimentally observed~\cite{nanobubbles_AFM}. 
Interesting objectives for future research are to clarify if the gas enrichment
constitutes a reservoir for nanobubbles and if the reduced diffusion of gas
perpendicular to the wall helps stabilizing them. 

What causes the dramatic gas enrichment? Energetically the system benefits
from gas-liquid interactions due to
$\epsilon_{gl}$, but gas particles in the bulk liquid occupy space
due to $\sigma_g$ which is unfavorable (reduction of liquid-liquid
interactions). Gases at the liquid interface, however,
reduce the energy with little disturbance of the liquid-liquid
interactions, and energetic contributions from $\epsilon_{lw}$, which are
diminished, are small for hydrophobic walls. According to this
explanation, the gas enrichment increases with
increasing $\epsilon_{gl}$ and $\sigma_g$ (for similar gas concentration in
the bulk liquid),
just as observed in the simulations. 

\begin{figure}[b]
\includegraphics[width=85mm]{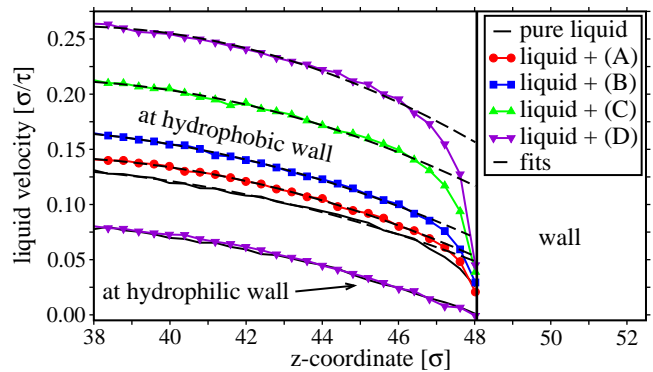}\vspace{0mm}
\caption{
\label{flow}
(color) Velocity profiles of the liquid close to the wall. The gas enrichment
at the hydrophobic wall clearly enhances the average velocity, contrary to the
hydrophilic case, where the gas has no influence. The dashed lines are
quadratic fits which are used to estimate the slip length (see text).
}
\end{figure}

To probe the effect of the gas on the slippage behavior, we apply a constant force
$f_y{=}2.27{\times}10^{-3}\epsilon_{ll}/\sigma$ in $y$-direction (parallel to
the wall). As usual only the velocity component perpendicular to the flow
($x$-component) is thermostated~\cite{bizonneEPJE,barrat} (Langevin
thermostat with $\tau _T{=}\tau$). The unforced systems after
their equilibration phase (initial configuration Fig.~\ref{setup}(c)
for $(B){-}(D)$) are further equilibrated ($1.2{\times}10^6dt$) while
applying $f_y$, and hereafter production runs ($10^6dt$) yield velocity
profiles shown in Fig.~\ref{flow}.  
The velocity in the liquid film at the hydrophilic wall is not altered by the
gas. Contrarily, at the hydrophobic wall the gas significantly changes the
velocity profiles, leading to an increase of the
average velocity. Estimates of the slip length $\lambda{=}|v_y/\partial_z
v_y|_{\rm wall}$~\cite{lauga} using the fits depicted in Fig.~\ref{flow} (dashed
lines) yield $\lambda{\approx}(3.7,3.4,4.5,7.0,7.9)\sigma$ for the pure
liquid and liquid in the presence of $(A){-}(D)$, respectively. Hence,
the presence of gas can significantly increase the slip length. 

In conclusion, our results support the experimental findings that gases
dissolved in liquids, although present only in low concentration in the bulk
liquid, can have a strong influence on the structure of the liquid-wall interface, due to
gas enrichment at hydrophobic walls. Future studies of phenomena associated
with the hydrophobic wall-liquid interface therefore must take dissolved gases
into account. This holds, e.g., for the appearance of nanoscale bubbles, the
study of slippage, and the breakage of nanofilms~\cite{jacobs}.

We thank the group of D.E.~Wolf for technical support, in particular
G.~Bartels and L.~Brendel. S.M.D.~acknowledges financial support (research
grant DA969/1-1) from the German
Research Foundation (DFG).


\begin{thebibliography}{99}
\bibitem{footnote2} Due to long diffusion times, the equilibrium gas
  enrichment lies between that obtained from the contrary initial
  configurations, Fig.~\ref{setup}(b,c).
\end{thebibliography}
\end{document}